\documentstyle[aps,prd,epsf]{revtex}
\begin{document}
%%%%%%%%%%%%%%%%%%%%%%%%%%%%%%%%%%%%%%%%%%%%%%%%%%%%%%%%%%%%%%%%%%%%%%
\title{Gravitational vacuum polarization I: \\
       Energy conditions in the Hartle--Hawking vacuum\\
       gr-qc/9604007}
\author{Matt Visser\cite{e-mail}}
\address{Physics Department, Washington University, St. Louis, 
         Missouri 63130-4899}
\date{1 April 1996}
\maketitle
\begin{abstract}

When a quantum field theory is constructed on a curved background
spacetime, the gravitationally induced vacuum polarization typically
induces a non-zero vacuum expectation value for the quantum
stress-energy tensor. It is well-known that this gravitational
vacuum polarization often violates the point-wise energy conditions
and sometimes violates the averaged energy conditions. In this
paper I begin a systematic attack on the question of where and by
how much the various energy conditions are violated.  To keep the
discussion manageable, I work in the test-field limit, and focus
on conformally coupled massless scalar fields in Schwarzschild
spacetime, using the Hartle--Hawking vacuum. The discussion invokes
a mixture of analytical and numerical techniques, and critically
compares the qualitative behaviour to be expected from the Page
approximation with that adduced from the numerical calculations of
Anderson, Hiscock, and Samuel.  I show that the various point-wise
energy conditions are violated in a series of onion-like layers
located between the unstable photon orbit and the event horizon,
the sequence of violations being DEC, WEC, and (NEC+SEC). Furthermore
the ANEC is violated for {\em some} of the null geodesics trapped
in this region. Having established the basic machinery in this
paper, the Boulware vacuum will be treated in a companion paper,
while studies of the Unruh vacuum should be straightforward, as
should extensions to non-conformal couplings, massive scalars, and
Reissner--Nordstr\"om geometries.

\end{abstract}

%%%%%%%%%%%%%%%%%%%%%%%%%%%%%%%%%%%%%%%%%%%%%%%%%%%%%%%%%%%%%%%%%%%%%%%%%%
%%%%%%%%%%%%%%%%%%%%%%%%%%%%%%%%%%%%%%%%%%%%%%%%%%%%%%%%%%%%%%%%%%%%%%%%%%
%%%\bibliographystyle{prsty}
%%%\bibliographystyle{unsrt}
%%%%%%%%%%%%%%%%%%%%%%%%%%%%%%%%%%%%%%%%%%%%%%%%%%%%%%%%%%%%%%%%%%%%%%%%%%
%%%\newtheorem{theorem}{Theorem}
%%%\newtheorem{definition}{Definition}
%%%%%%%%%%%%%%%%%%%%%%%%%%%%%%%%%%%%%%%%%%%%%%%%%%%%%%%%%%%%%%%%%%%%%%%%%%
\widetext
\section{INTRODUCTION}
%%%%%%%%%%%%%%%%%%%%%%%%%%%%%%%%%%%%%%%%%%%%%%%%%%%%%%%%%%%%%%%%%%%%%%%%%%

When a quantum field theory is constructed on a curved background
spacetime, the gravitationally induced vacuum polarization typically
induces a non-zero vacuum expectation value for the stress-energy
tensor~\cite{deWitt75,Centenary,Birrell-Davies,Three-Hundred,Fulling,Visser}.
It should be pointed out that fully self consistent calculations
are horrendously difficult; consequently most calculations in the
literature are carried out in the test-field limit. (Wherein the
vacuum polarization is {\em not} permitted to back-react on the
geometry via the Einstein field equations.) Even in the test-field
limit, most known results are obtained by numerical (rather than
analytic) computations.

{\em Perturbative} self-consistent solutions around flat spacetime
have recently been investigated by Flanagan and Wald~\cite{Flanagan-Wald},
building on earlier work of Horowitz~\cite{Horowitz80}, and Horowitz
and Wald~\cite{Horowitz-Wald}.

Another type of energy condition, based on the ``quantum inequalities'',
is investigated for a Schwarzschild background in recent papers by
Ford and Roman~\cite{Ford-Roman93,Ford-Roman96}.

Ford and Roman have also discussed the ANEC and AWEC in (1+1)--dimensional
and (3+1)--dimensional evaporating black holes (Unruh
vacuum)~\cite{Ford-Roman96}, and for the (1+1)--dimensional Boulware
vacuum.  In contrast, the present paper works strictly in (3+1)
dimensions and treats the equilibrium Hartle--Hawking vacuum state.

Now we know, on rather general grounds (the existence of the Hawking
radiation effect and concomitant violation of the {\em classical}
area increase theorem of black hole dynamics) that at least some
of the energy conditions must be violated at or near the event
horizon of a black hole.  In this paper I shall explore these issues
in a little more detail. I shall continue to work in the test-field
limit: I discuss the effect of the gravitational vacuum polarization
on various energy conditions. In particular, I discuss the point-wise
null, weak, strong, and dominant energy conditions (NEC, WEC, SEC,
and DEC), the averaged null energy condition (ANEC), and shall
furthermore introduce and discuss several new, potentially interesting,
energy conditions (PNEC, Scri--NEC, and Scri--ANEC).

For the purposes of this paper, I restrict attention to the most
well-studied curved-space quantum field theory: the conformally-coupled
massless scalar field on Schwarzschild spacetime, in the Hartle--Hawking
vacuum. For this geometry and vacuum state one has both (1) a useful
analytic approximation to the gravitational polarization---Page's
approximation~\cite{Page82}, and (2) numerical estimates of the
vacuum polarization---see the numerical calculations of
Howard~\cite{Howard}, Howard and Candelas~\cite{Howard-Candelas},
and Anderson, Hiscock, and Samuel~\cite{AHS,AHS-2,Anderson-Private}.)

Having established the basic machinery in this paper, a study of
the Boulware vacuum will be presented in a companion paper, while
further studies of the Unruh vacua should be straightforward,
although tedious. (For the current state of affairs in the Unruh
vacuum, see~\cite{Ford-Roman96}.) Similarly, extensions to
non-conformal couplings, massive scalars, and Reissner--Nordstr\"om
geometries should be straightforward.

%%%%%%%%%%%%%%%%%%%%%%%%%%%%%%%%%%%%%%%%%%%%%%%%%%%%%%%%%%%%%%%%%%%%%%%%
\section{Null energy conditions}
%%%%%%%%%%%%%%%%%%%%%%%%%%%%%%%%%%%%%%%%%%%%%%%%%%%%%%%%%%%%%%%%%%%%%%%%

To set the stage, recall some basic definitions ---\\
%%%\begin{definition}[NEC]\hfil\break
{\bf Definition [NEC]}\hfil\break
{\em
The null energy condition is said
to hold at a point  $p$ if for  all null vectors $k^\mu$
\begin{equation}
T_{\mu\nu} \;  k^\mu k^\nu \geq 0. 
\end{equation}
}
%%%\end{definition}

%%%\begin{definition}[ANEC]\hfil\break
\noindent{\bf Definition [ANEC]}\hfil\break
{\em
The averaged null energy condition is said
to hold on a null curve $\gamma$ if
\begin{equation}
\int_\gamma T_{\mu\nu} \;  k^\mu k^\nu \; d\lambda \geq 0. 
\end{equation}
}
%%%\end{definition}

Here $\lambda$ is a generalized affine parameter
(see~\cite[pp. 259, 278, 291]{Hawking-Ellis}) for the null curve,
the tangent vector being denoted by $k^\mu$.

Note that in almost all applications it suffices to consider null
geodesics, rather than generic null curves. However, isolated
discussions of the ANEC on non--geodesic null curves do occur in
the literature~\cite{Klinkhammer91}---the extended definition
presented here allows one to compare and contrast the present
results with other calculations.

As a practical matter I shall often replace the word ``inextendible''
by the phrase ``inextendible past the event horizon''. This is a
purely pragmatic decision based on a lack of numerical data inside
the event horizon, coupled with the general feeling in the community
that Page's analytic approximation will become progressively worse
as one approaches the central singularity. (Thus my studies of the
ANEC can more precisely be referred to as studies of a
``Truncated--ANEC''.)

In an earlier publication~\cite{Visser95}, (see also~\cite{Visser}),
I derived a no-go theorem for the ANEC: I showed that the ANEC is
violated (in the test-field limit) whenever the background spacetime
has a non-zero {\em scale anomaly}. Specifically, I proved the\\
%%%\begin{theorem}[ANEC no-go]
{\bf ANEC no-go theorem}\hfil\break
{\em 
In $(3+1)$--dimensional spacetime, $({\cal M}, g)$, for any
conformally coupled massless quantum field, in any conformal quantum state:
If
\begin{equation}
Z^\mu{}_\nu \equiv 
\left[\nabla_\alpha \nabla^\beta + {1\over2} R^\alpha{}_\beta \right] 
C^{\alpha\mu}{}_{\beta\nu}  \neq 0
\end{equation}
then, holding the renormalization scale $\mu$ fixed, it is possible
to find a rescaled metric $\bar g = \Omega^2 g$ such that the ANEC
is violated on the spacetime $({\cal M}, \bar g)$.
}
%%%\end{theorem}

This no-go theorem, since it depends on a special case of the
quantum field theoretic (conformal) anomaly, is known to be
independent of the choice of vacuum state. (One can easily generalize
this argument to include non-conformally coupled fields,
see~\cite{Flanagan-Wald}.)

Unfortunately, a brief calculation suffices to show that the tensor
$Z^\mu{}_\nu$ vanishes on Schwarzschild spacetime. Indeed it vanishes
on any Ricci-flat spacetime (or indeed on any spacetime conformal
to an Einstein spacetime). Thus Schwarzschild spacetime avoids the
no--go theorem enunciated above, and it becomes a matter of explicit
checking to see whether or not the ANEC is satisfied or violated
on this geometry. Note that in doing this explicit checking we will
have to be cognizant of our particular choice of vacuum state, and
indeed our results will also depend on the particular class of null
geodesics under consideration.

In performing the explicit checks alluded to above, I have found
it useful to define several new energy conditions that are in some
sense intermediate between the NEC and ANEC:

%%%\begin{definition}[PNEC]\hfil\break
\noindent{\bf Definition [PNEC]}\hfil\break
{\em
The partial null energy condition will be said to hold  at a point
$p$ for a set of null tangent vectors  $P(p)\subset T_p$ if for 
all null vectors $k^\mu \in P(p)$
\begin{equation}
T_{\mu\nu} \;  k^\mu k^\nu \geq 0. 
\end{equation}
}
%%%\end{definition}

Now, if one does not place any further constraints on the set $P(p)$
the resulting definition, while general, is not particularly useful.
(It is best used as a diagnostic tool to probe interesting regions
of the tangent bundle.) A suitable restriction is:

%%%\begin{definition}[Scri--NEC]\hfil\break
\noindent{\bf Definition [Scri--NEC]}\hfil\break
{\em
The asymptotic null energy condition will be said to hold on an
asymptotically flat spacetime if for every point $p$ of the spacetime,
and all null tangent vectors $k^\mu$ such that the associated null
geodesic through $p$ either (1) escapes to future null infinity
{\em (Scri$^+$)}, or (2) arrives from past null infinity {\em
(Scri$^-$)}, one has:
\begin{equation}
T_{\mu\nu} \;  k^\mu k^\nu \geq 0. 
\end{equation}
}
%%%\end{definition}

(This definition basically says that one should not worry too much
if violations of the NEC occur only when one looks along null
geodesics that never make it out to null infinity.)

We shall soon see that in the Hartle--Hawking vacuum the gravitational
vacuum polarization of a conformally-coupled massless scalar field residing
on the Schwarzschild background geometry everywhere satisfies this
Scri--NEC energy condition, though it does not everywhere satisfy
the NEC. We shall also use the PNEC to investigate the regions in
phase space (the tangent bundle) where suitable positivity conditions
hold.

Finally, for completeness, I enunciate\\
%%%\begin{definition}[Scri--ANEC]\hfil\break
\noindent{\bf Definition [Scri--ANEC]}\hfil\break
{\em
The~asymptotic~averaged null energy condition will be said to hold
on an asymptotically flat spacetime if every  inextendible null
curve which either (1) escapes to future null infinity {\em(Scri$^+$)},
or (2) arrives from past null infinity {\em (Scri$^-$)}, satisfies the
ANEC:
\begin{equation}
\int_\gamma T_{\mu\nu} \;  k^\mu k^\nu \; d\lambda \geq 0. 
\end{equation}
}
%%%\end{definition}

(This definition basically says that one should not worry too much
if violations of the ANEC occur only along null geodesics that
never make it out to null infinity. After all, null infinity is
where you want to work to establish results such as the
Penrose--Sorkin--Woolgar positive mass
theorem~\cite{Penrose-Sorkin-Woolgar} or the Friedman--Schleich--Witt
topological censorship theorem~\cite{Topological-censorship}.)

From the previous comment regarding Scri--NEC it automatically
follows that, in the Hartle--Hawking vacuum, the gravitational
vacuum polarization of a conformally coupled scalar field residing
on the Schwarzschild background geometry satisfies this Scri--ANEC.
We shall also see that not all null geodesics satisfy the ANEC
itself.

%%%%%%%%%%%%%%%%%%%%%%%%%%%%%%%%%%%%%%%%%%%%%%%%%%%%%%%%%%%%%%%%%%%%%%%%%%%%%%
\section{Vacuum polarization in Schwarzschild spacetime:
         Hartle--Hawking vacuum}
%%%%%%%%%%%%%%%%%%%%%%%%%%%%%%%%%%%%%%%%%%%%%%%%%%%%%%%%%%%%%%%%%%%%%%%%%%%%%%

In the particular case of conformally coupled scalar fields residing
on Schwarzschild spacetime one has a lot of information about the
vacuum polarization. By spherical symmetry one knows that
\begin{equation}
\langle H | T^{\hat\mu}{}_{\hat\nu} | H \rangle \equiv 
\left[ \matrix{-\rho&0&0&0\cr
               0&-\tau&0&0\cr
               0&0&p&0\cr
	       0&0&0&p\cr } \right].
\end{equation}
Where $\rho$, $\tau$ and $p$ are functions of $r$, $M$ and $\hbar$.
Note that I set $G\equiv1$, and choose to work in a local-Lorentz
basis attached to the fiducial static observers (FIDOS).

Page's analytic approximation gives~\cite{Page82,Howard} a polynomial
approximation to the stress--energy tensor:
\widetext
\begin{eqnarray}
\rho(r) &=&  +3 p_\infty 
\left[
1 + 2\left({2M\over r}\right) + 3\left({2M\over r}\right)^2 
+ 4\left({2M\over r}\right)^3 + 5\left({2M\over r}\right)^4 
+ 6\left({2M\over r}\right)^5 -33\left({2M\over r}\right)^6
\right],
\\
%%%%%%%%%%%%%%%%%%%%%%%%%%%%%%%%%%%%%%%%%%%%%%%%%%%%%%%%%%%%%%%%%%%%%%%%%%%%%%
%             2       3       4       5       6
%    6 R   9 R    12 R    15 R    18 R    99 R
%3 + --- + ---- + ----- + ----- + ----- - -----
%     r      2      3       4       5       6
%           r      r       r       r       r
%%%%%%%%%%%%%%%%%%%%%%%%%%%%%%%%%%%%%%%%%%%%%%%%%%%%%%%%%%%%%%%%%%%%%%%%%%%%%%  
\tau(r) &=& - p_\infty 
\left[ 
1 + 2\left({2M\over r}\right) + 3\left({2M\over r}\right)^2 
+ 4\left({2M\over r}\right)^3 + 5\left({2M\over r}\right)^4 
+ 6\left({2M\over r}\right)^5 +15\left({2M\over r}\right)^6
\right],
\\
%%%%%%%%%%%%%%%%%%%%%%%%%%%%%%%%%%%%%%%%%%%%%%%%%%%%%%%%%%%%%%%%%%%%%%%%%%%%%%
%             2      3      4      5       6
%    2 R   3 R    4 R    5 R    6 R    15 R
%1 + --- + ---- + ---- + ---- + ---- + -----
%     r      2      3      4      5      6
%           r      r      r      r      r
%%%%%%%%%%%%%%%%%%%%%%%%%%%%%%%%%%%%%%%%%%%%%%%%%%%%%%%%%%%%%%%%%%%%%%%%%%%%%%
p(r)    &=& +p_\infty 
\left[ 
1 + 2\left({2M\over r}\right) + 3\left({2M\over r}\right)^2 
+ 4\left({2M\over r}\right)^3 + 5\left({2M\over r}\right)^4 
+ 6\left({2M\over r}\right)^5 - 9\left({2M\over r}\right)^6
\right],
\\
%%%%%%%%%%%%%%%%%%%%%%%%%%%%%%%%%%%%%%%%%%%%%%%%%%%%%%%%%%%%%%%%%%%%%%%%%%%%%%
%             2      3      4      5      6
%    2 R   3 R    4 R    5 R    6 R    9 R
%1 + --- + ---- + ---- + ---- + ---- - ----
%     r      2      3      4      5      6
%           r      r      r      r      r
%%%%%%%%%%%%%%%%%%%%%%%%%%%%%%%%%%%%%%%%%%%%%%%%%%%%%%%%%%%%%%%%%%%%%%%%%%%%%%
       &=& +p_\infty 
\left[ 
1 + 2\left({2M\over r}\right) + 3\left({2M\over r}\right)^2 
\right]
\left[
1+  4\left({2M\over r}\right)^3 - 3\left({2M\over r}\right)^4
\right]. 
\end{eqnarray}
%%%\narrowtext
Here I have defined a constant (to be interpreted as the pressure
at spatial infinity) by
\begin{equation}
p_\infty \equiv {\hbar\over 90 (16\pi)^2 (2M)^4}. 
\end{equation}
Note that I have explicitly expanded the functions given in
references~\cite{Page82,Howard} as polynomials in $2M/r$ to exhibit
the fact that all components of the stress--energy are well behaved
at the horizon. It is perhaps somewhat surprising that the rather
messy rational {\em quotients} of polynomials exhibited
in~\cite{Page82,Howard} reduce to such relatively nice compact
expressions. These formulae have also been checked for consistency
with the Brown--Ottewill extensions of the original Page
approximation~\cite[see esp. p 2517]{Brown-Ottewill}. See also
Elster~\cite{Elster83}. If one prefers, an alternative form for
the Page approximation is
\widetext
\begin{equation}
\langle H | T^{\hat\mu}{}_{\hat\nu} | H \rangle \equiv 
p_\infty \;
\left\{ P(r)
\left[ \matrix{-3&0&0&0\cr
               0&+1&0&0\cr
               0&0&+1&0\cr
	       0&0&0&+1\cr } \right]
+ Q(r) 	   
\left[ \matrix{+3&0&0&0\cr
               0&+1&0&0\cr
               0&0&0&0\cr
	       0&0&0&0\cr } \right]
\right\}.
\end{equation}
The polynomials $P(r)$ and $Q(r)$ being 
\begin{equation}
P(r) = 
1 +2\left({2M\over r}\right) +3\left({2M\over r}\right)^2 
+4\left({2M\over r}\right)^3 +5\left({2M\over r}\right)^4 
+6\left({2M\over r}\right)^5 - 9\left({2M\over r}\right)^6,
\end{equation}
%%%\narrowtext
and 
\begin{equation}
Q(r) = 24 \; \left({2M\over r}\right)^6.
\end{equation}

The trace of the stress--energy tensor is given by
\begin{equation}
\langle H | T^{\hat\mu}{}_{\hat\mu} | H \rangle \equiv 
96 \; p_\infty \;
\left({2M\over r}\right)^6. 	 
\end{equation}
This result, because it is simply a restatement of the conformal
anomaly, is known to be exact.

At the event horizon, Page's analytic approximation
gives~\cite{Page82,Howard}
\begin{equation}
\langle H | T^{\hat\mu}{}_{\hat\nu} | H \rangle \equiv 
p_\infty 
\left[ \matrix{+36&0&0&0\cr
               0&+36&0&0\cr
               0&0&+12&0\cr
	       0&0&0&+12\cr } \right].
\end{equation}

The explicit numerical calculations of Howard~\cite{Howard}, Howard
and Candelas~\cite{Howard-Candelas}, and Anderson, Hiscock, and
Samuel~\cite{AHS,AHS-2,Anderson-Private} show that Page's analytic
approximation is reasonably good---the worst deviations occur in
the immediate vicinity of the event horizon where the
Anderson--Hiscock--Samuel numerical analysis  gives
\widetext
\begin{equation}
\langle H | T^{\hat\mu}{}_{\hat\nu} | H \rangle \approx  
p_\infty
\left[ \matrix{+37.7403&0&0&0\cr
               0&+37.7403&0&0\cr
               0&0&+10.259&0\cr
	       0&0&0&+10.259\cr } \right].
\end{equation}
%%%\narrowtext
A convenient table of numerical results, covering the range $r=2M$
to $r=6.8M$ may be found on page 2541 of reference~\cite{Howard}.
(Near the event horizon, the Howard--Candelas data is believed to
be accurate to two significant digits. The accuracy is expected to
improve with increasing $r$, as the numerical results converge upon
the Page approximation.) 

Additionally, I have obtained via private communication the more
intensive numerical data of
Anderson--Hiscock--Samuel~\cite{AHS,AHS-2,Anderson-Private}, which
covers the range $r=2M$ to $r=5M$ at much higher resolution. (Near
the event horizon, the Anderson--Hiscock--Samuel data is believed
to be accurate to three significant digits.) I have collated the
data and used it to construct suitable interpolating functions to
be used for the numerical aspects of the following discussion.

To construct the interpolating functions I use the
Anderson--Hiscock--Samuel data in the range $r=2M$ to $r=5M$, and
the Howard--Candelas data in the range $r=5M$ to $r=6.8M$, augmented
by the fact that we know the exact result at $r=\infty$.  Using
Mathematica, I have fit the data using third-order interpolating
polynomials in the variable $z=2M/r$. (Note that $z$ conveniently
runs from $z=1$ at the horizon to $z=0$ at spatial infinity; $z$
is also {\em the} natural variable to use when numerically analyzing
the Page approximation.)

Of course, there are may other papers extant which calculate the
stress-energy tensor in somewhat different configurations. See,
for example,
\cite{Page82,AHS,AHS-2,Brown-Ottewill,Elster83,Brown-Ottewill-Page,%
Candelas-Howard,Frolov-Thorne,Fulling77,CCH,Jensen-Ottewill,Zelnikov,%
Elster84,JLO91,JLO92}. More recently, in a very interesting
development, Matyjasek~\cite{Matyjasek} has developed a curve-fitting
analysis that fits the numerical data to high accuracy. In this
paper I prefer to work directly with the numerical data.  In this
paper I am only presenting the minimum requirements for the particular
job at hand and it is clear that with additional effort more
information can be extracted by mining this additional vein of
data.

%%%%%%%%%%%%%%%%%%%%%%%%%%%%%%%%%%%%%%%%%%%%%%%%%%%%%%%%%%%%%%%%%%%%%%%%%%%%%%
%%%% To further simplify notation, I shall henceforth set $G=1$. 
%%%% For future reference I explicitly state the Page approximation 
%%%% in the form
%%%%%%%%%%%%%%%%%%%%%%%%%%%%%%%%%%%%%%%%%%%%%%%%%%%%%%%%%%%%%%%%%%%%%%%%%%%%%%
\section{Null energy condition}
%%%%%%%%%%%%%%%%%%%%%%%%%%%%%%%%%%%%%%%%%%%%%%%%%%%%%%%%%%%%%%%%%%%%%%%%%%%%%%
\subsection{Outside the horizon:}
%%%%%%%%%%%%%%%%%%%%%%%%%%%%%%%%%%%%%%%%%%%%%%%%%%%%%%%%%%%%%%%%%%%%%%%%%%%%%%

Outside the event horizon, the NEC reduces to the pair of constraints
\begin{equation}
\rho(r) - \tau(r) \geq 0? \qquad \rho(r) + p(r) \geq 0?
\end{equation}
Page's approximation yields
\widetext
\begin{equation}
\rho(r) - \tau(r) = 4 p_\infty
\left(1-{2M \over r}\right) 
\left[
1+ 3\left({2M\over r}\right)    +6\left({2M\over r}\right)^2 
+ 10\left({2M\over r}\right)^3 +15\left({2M\over r}\right)^4
+ 21\left({2M\over r}\right)^5
\right].
\end{equation}
%%%\narrowtext
%%%%%%%%%%%%%%%%%%%%%%%%%%%%%%%%%%%%%%%%%%%%%%%%%%%%%%%%%%%%%%%%%%%%%%%%%%%%%%
%                   5        4         3  2       2  3        4    5
%4 (-2 m + r) (672 m  + 240 m  r + 80 m  r  + 24 m  r  + 6 m r  + r )
%--------------------------------------------------------------------
%                                  6
%                                 r
%%%%%%%%%%%%%%%%%%%%%%%%%%%%%%%%%%%%%%%%%%%%%%%%%%%%%%%%%%%%%%%%%%%%%%%%%%%%%%
(Note the factorization!) Consequently $\rho(r) - \tau(r)$ is always
explicitly positive outside the horizon. Furthermore 
\widetext
\begin{equation}
\rho(r) + p(r) =  4 p_\infty
\left[
1+2\left({2M\over r}\right)   + 3 \left({2M\over r}\right)^2
+ 4\left({2M\over r}\right)^3 + 5 \left({2M\over r}\right)^4
+ 6\left({2M\over r}\right)^5 - 27\left({2M\over r}\right)^6
\right].
\end{equation}
%%%\narrowtext
%%%%%%%%%%%%%%%%%%%%%%%%%%%%%%%%%%%%%%%%%%%%%%%%%%%%%%%%%%%%%%%%%%%%%%%%%%%%%%
%                2      3      4      5       6
%       2 R   3 R    4 R    5 R    6 R    27 R
% 4( 1 + --- + ---- + ---- + ---- + ---- - ----- )
%         r      2      3      4      5      6
%               r      r      r      r      r
%%%%%%%%%%%%%%%%%%%%%%%%%%%%%%%%%%%%%%%%%%%%%%%%%%%%%%%%%%%%%%%%%%%%%%%%%%%%%%
Numerically finding the roots of this sixth order polynomial shows
that $\rho(r) + p(r)$ is negative from $r=2M$ to $r\approx 2.18994M$.

Thus the Page approximation suggests that the null energy condition
is violated in the range $r\in[2M,2.18994M]$.

Turning to the numerical data, one easily
verifies that that outside the horizon $\rho(r) -\tau(r) >0$, while
$\rho = \tau$ at the horizon. This is in complete agreement with
Page's analytic approximation.

On the other hand, inspection of the numerical data indicates
that $\rho+p < 0$ for  $r \lesssim 2.298\, M$. (This number was
obtained by fitting third-order interpolating polynomials to
the numeric data; and then numerically finding the root.)
This is qualitatively, though not quantitatively in agreement with
Page's analytic approximation.

Thus the numerical data indicate that the null energy condition is
violated in the range $r\in[2M,2.298M]$.

%%%%%%%%%%%%%%%%%%%%%%%%%%%%%%%%%%%%%%%%%%%%%%%%%%%%%%%%%%%%%%%%%%%%%%%%%%%%%%
\subsection{Inside the horizon:}
%%%%%%%%%%%%%%%%%%%%%%%%%%%%%%%%%%%%%%%%%%%%%%%%%%%%%%%%%%%%%%%%%%%%%%%%%%%%%%

Inside the event horizon, the radial coordinate becomes timelike,
and the roles played by $\rho(r)$ and $\tau(r)$ are interchanged.
The NEC reduces to the pair of constraints
\begin{equation}
\tau(r) -\rho(r)\geq 0? \qquad \tau(r) + p(r) \geq 0?
\end{equation}

Unfortunately, inside the horizon I am reduced to reliance upon
the Page approximation by a total lack of numerical data. Now there
are some subtleties associated with extrapolating the Page
approximation inside the event horizon. The existence of a Killing
vector that points in the $t$ direction is absolutely critical,
(even if the $t$ direction is no longer timelike).  Thus if you
wish to extrapolate the Page approximation inside the event horizon
it is necessary to limit attention to an {\em eternal} black hole
(that is, the maximally extended Kruskal--Szekeres manifold).

I should remind the reader that in the Hartle--Hawking vacuum the
stress-energy is at least known to be regular at the horizon, so
the Page approximation should not be too far wrong just inside the
event horizon of an eternal black hole. On the other hand,
one does expect the Page approximation to get progressively worse
as one moves further in toward the singularity, so one should
probably not take these results too seriously once one is far inside
the horizon.

I should especially warn the reader against extrapolating the Page
approximation to the interior of an astrophysical black hole: we
really do not expect the interior of an astrophysical black hole
to have the same Killing vectors as the exterior, but instead expect
the interior to be fully dynamical. (In addition we do not expect
an astrophysical black hole to even to be in the Hartle--Hawking
quantum state.)

The present calculations inside the horizon are strictly limited
to eternal black holes and are exhibited merely because they are
do-able, and because they may be suggestive of the actual situation.

Inspection of Page's analytic approximation indicates that $ \tau(r)
-\rho(r) >0$ for all $r<2M$, so this part of the NEC is satisfied
inside the event horizon. On the other hand
\begin{equation}
\tau(r) +p(r) = - 24 p_\infty   \left({2M\over r}\right)^6,
\end{equation}
which is everywhere negative, (both inside and outside the horizon).
Thus the Page approximation suggests that the null energy condition
is violated throughout the interior of the black hole.

This is enough to inform us that, (assuming the reliability of the
Page approximation in this matter), the weak, strong, and dominant
energy conditions are also violated throughout the entire interior
of the black hole.

%%%%%%%%%%%%%%%%%%%%%%%%%%%%%%%%%%%%%%%%%%%%%%%%%%%%%%%%%%%%%%%%%%%%%%%%%%%%%%
\section{Weak energy condition}
%%%%%%%%%%%%%%%%%%%%%%%%%%%%%%%%%%%%%%%%%%%%%%%%%%%%%%%%%%%%%%%%%%%%%%%%%%%%%%

%%%%%%%%%%%%%%%%%%%%%%%%%%%%%%%%%%%%%%%%%%%%%%%%%%%%%%%%%%%%%%%%%%%%%%%%%%%%%%
\subsection{Outside the horizon:}
%%%%%%%%%%%%%%%%%%%%%%%%%%%%%%%%%%%%%%%%%%%%%%%%%%%%%%%%%%%%%%%%%%%%%%%%%%%%%%

Outside the horizon, the weak energy condition is equivalent to
the three constraints
\begin{eqnarray}
&&\rho(r) \geq 0?
\nonumber\\ 
&&\rho(r) - \tau(r) \geq 0? 
\nonumber\\ 
&&\rho(r) + p(r) \geq 0?
\end{eqnarray}
The last two constraints have already been discussed---they are
simply the null energy condition. In the Page approximation, finding
the roots of the relevant sixth-order polynomial shows that the
density constraint is violated for $r < 2.3468 M$. Therefore the
weak energy condition is violated in the range $r\in[2M,2.3468M]$.

In contrast, the numerical data indicate that violations of
the weak energy condition extend to the region $r\in[2M,2.438M]$.

%%%%%%%%%%%%%%%%%%%%%%%%%%%%%%%%%%%%%%%%%%%%%%%%%%%%%%%%%%%%%%%%%%%%%%%%%%%%%%
\subsection{Inside the horizon:}
%%%%%%%%%%%%%%%%%%%%%%%%%%%%%%%%%%%%%%%%%%%%%%%%%%%%%%%%%%%%%%%%%%%%%%%%%%%%%%

Inside the horizon, the weak energy condition is equivalent to the
three constraints
\begin{eqnarray}
&&\tau(r) \geq 0?
\nonumber\\
&&\tau(r) -\rho(r)\geq 0?
\nonumber\\
&&\tau(r) + p(r) \geq 0?
\end{eqnarray}
The last two constraints have already been discussed (when dealing
with the NEC). The violation of the NEC is already sufficient to
tell us that the WEC is violated everywhere inside the horizon,
but we can add in passing that the Page approximation additionally
implies that $\tau(r)$ is everywhere negative inside (and outside)
the horizon.

The Page approximation suggests that the weak energy condition is
violated throughout the interior of the black hole.

%%%%%%%%%%%%%%%%%%%%%%%%%%%%%%%%%%%%%%%%%%%%%%%%%%%%%%%%%%%%%%%%%%%%%%%%%%%%%%
\section{Strong energy condition}
%%%%%%%%%%%%%%%%%%%%%%%%%%%%%%%%%%%%%%%%%%%%%%%%%%%%%%%%%%%%%%%%%%%%%%%%%%%%%%

%%%%%%%%%%%%%%%%%%%%%%%%%%%%%%%%%%%%%%%%%%%%%%%%%%%%%%%%%%%%%%%%%%%%%%%%%%%%%%
\subsection{Outside the horizon:}
%%%%%%%%%%%%%%%%%%%%%%%%%%%%%%%%%%%%%%%%%%%%%%%%%%%%%%%%%%%%%%%%%%%%%%%%%%%%%%

Outside the horizon, the strong energy condition is equivalent
to the three constraints
\begin{eqnarray}
&&\rho(r) - \tau(r)     \geq 0?
\nonumber\\
&&\rho(r)+p(r)          \geq 0?
\nonumber\\
&&\rho(r)-\tau(r)+2p(r) \geq 0?
\end{eqnarray}
We have already looked at the first two constraints when discussing
the null energy condition. The third condition is always satisfied
outside the horizon (since both $\rho(r)-\tau(r)>0$, and $p(r)>0$
in this region). This is true both in the Page approximation, and
by appeal to the  numerical data. Thus the strong energy
condition is violated in the same region as the null energy condition.

%%%%%%%%%%%%%%%%%%%%%%%%%%%%%%%%%%%%%%%%%%%%%%%%%%%%%%%%%%%%%%%%%%%%%%%%%%%%%%
\subsection{Inside the horizon:}
%%%%%%%%%%%%%%%%%%%%%%%%%%%%%%%%%%%%%%%%%%%%%%%%%%%%%%%%%%%%%%%%%%%%%%%%%%%%%%

Inside the horizon, the strong energy condition is equivalent
to the three constraints
\begin{eqnarray}
&&\tau(r) - \rho(r)         \geq 0?
\nonumber\\ 
&&\tau(r) + p(r)            \geq 0?
\nonumber\\
&&\tau(r) - \rho(r) + 2p(r) \geq 0?
\end{eqnarray}
We have already looked at the first two constraints when discussing
the null energy condition, and thereby know that (the Page
approximation suggests that) the strong energy condition is violated
throughout the entire interior of the black hole.

For completeness I point out that the first condition is satisfied
throughout the interior, the second of these conditions is violated
throughout the interior, while the third condition also fails
throughout the interior.

%%%%%%%%%%%%%%%%%%%%%%%%%%%%%%%%%%%%%%%%%%%%%%%%%%%%%%%%%%%%%%%%%%%%%%%%%%%%%%
\section{Dominant energy condition}
%%%%%%%%%%%%%%%%%%%%%%%%%%%%%%%%%%%%%%%%%%%%%%%%%%%%%%%%%%%%%%%%%%%%%%%%%%%%%%

%%%%%%%%%%%%%%%%%%%%%%%%%%%%%%%%%%%%%%%%%%%%%%%%%%%%%%%%%%%%%%%%%%%%%%%%%%%%%%
\subsection{Outside the horizon:}
%%%%%%%%%%%%%%%%%%%%%%%%%%%%%%%%%%%%%%%%%%%%%%%%%%%%%%%%%%%%%%%%%%%%%%%%%%%%%%

Outside the horizon, the dominant energy condition is equivalent
to the three constraints
\begin{eqnarray}
&&\rho(r) \geq 0? 
\nonumber\\
&&\tau(r) \in [-\rho(r),+\rho(r)]?
\nonumber\\
&&p(r)\in [-\rho(r),+\rho(r)]?
\end{eqnarray}
We can rephrase this as
\begin{equation}
\rho(r) \geq 0? \qquad 
\rho(r) \pm\tau(r) \geq 0? \qquad 
\rho(r) \pm p(r) \geq 0?
\end{equation}

Using the Page approximation, and restricting attention to the
range $[2M,\infty]$, one has:
\begin{eqnarray}
&\rho(r)<0           &\qquad     r\in[2M,2.3468M];   \nonumber\\
&\rho(r)-\tau(r)< 0  &\qquad     r\in\emptyset;      \nonumber\\
&\rho(r)+\tau(r) < 0 &\qquad     r\in[2M,2.77256M];  \nonumber\\
&\rho(r)-p(r) < 0    &\qquad     r\in[2M,2.58512M];  \nonumber\\
&\rho(r)+p(r) < 0    &\qquad     r\in[2M,2.18994M].
\end{eqnarray}
Pulling this all together, the Page approximation suggests that
dominant energy condition fails in the region $r\in[2M,2.77256M]$.

The numerical data implies quantitative though not qualitative
modifications. In the  range $[2M,\infty]$ one finds: 
\begin{eqnarray}
&\rho(r)<0           &\qquad     r\in[2M,2.438M];   \nonumber\\
&\rho(r)-\tau(r)< 0  &\qquad     r\in\emptyset;     \nonumber\\
&\rho(r)+\tau(r) < 0 &\qquad     r\in[2M,2.992M];   \nonumber\\
&\rho(r)-p(r) < 0    &\qquad     r\in[2M,2.628M];   \nonumber\\
&\rho(r)+p(r) < 0    &\qquad     r\in[2M,2.298M].
\end{eqnarray}
Pulling this all together, the numerical data indicate that the
dominant energy condition fails in the region $r\in[2M,2.992M]$.
Note that this is suspiciously close to $r=3M$---the unstable
circular photon orbit---and {\em might} be trying to tell us something.

If one actually calculates $(\rho+\tau)/(|\rho|+|\tau|)$
at $r=3M$ one gets $6.02\times10^{-3}$---given the expected
three-significant-digit numerical reliability of the data this is
(questionably) compatible with zero. If one takes this suggestion
seriously, it would imply a hidden (accidental?) symmetry in the
stress tensor at $r=3M$:
\begin{equation}
\left\langle H | T^{\hat\mu\hat\nu} | H \rangle\right|_{r=3M} \propto
\left[ \matrix{a&0&0&0\cr
               0&a&0&0\cr
               0&0&b&0\cr
	       0&0&0&b\cr } \right]?
\end{equation}
It is quite possible however, that this is purely a numerical accident.

Finally, I point out that this conjectured property certainly does
not survive the introduction of non-conformal coupling, nor is
there any particular reason to expect it to survive the introduction
of non-zero rest mass. This conjecture also most definitely does
not hold in the Boulware or Unruh vacuum
states~\cite{JLO91,JLO92,Ford-Roman96}.

%%%%%%%%%%%%%%%%%%%%%%%%%%%%%%%%%%%%%%%%%%%%%%%%%%%%%%%%%%%%%%%%%%%%%%%%%%%%%%
\subsection{Inside the horizon:}
%%%%%%%%%%%%%%%%%%%%%%%%%%%%%%%%%%%%%%%%%%%%%%%%%%%%%%%%%%%%%%%%%%%%%%%%%%%%%%

Inside the horizon, the dominant energy condition is equivalent to
the three constraints
\begin{eqnarray}
&&\tau(r) \geq 0?
\nonumber\\
&&\rho(r) \in [-\tau(r),+\tau(r)]? 
\nonumber\\ 
&&p(r)\in [-\tau(r),+\tau(r)]?
\end{eqnarray}
We can rephrase this as
\begin{equation}
-\tau(r) \geq 0? \qquad 
-\tau\pm\rho(r) \geq 0? \qquad 
-\tau\pm p(r) \geq 0?
\end{equation}
We already know that $\tau(r)< 0$ inside the horizon (in fact
everywhere).  So, provided the Page approximation is not misleading
in this regard,  the dominant energy condition is violated throughout
the interior of the black hole. No additional information comes
from the other conditions.

%%%%%%%%%%%%%%%%%%%%%%%%%%%%%%%%%%%%%%%%%%%%%%%%%%%%%%%%%%%%%%%%%%%%%%%%%%%%%%
\section{Partial null energy condition}
%%%%%%%%%%%%%%%%%%%%%%%%%%%%%%%%%%%%%%%%%%%%%%%%%%%%%%%%%%%%%%%%%%%%%%%%%%%%%%

%%%%%%%%%%%%%%%%%%%%%%%%%%%%%%%%%%%%%%%%%%%%%%%%%%%%%%%%%%%%%%%%%%%%%%%%%%%%%%
\subsection{Outside the horizon:}
%%%%%%%%%%%%%%%%%%%%%%%%%%%%%%%%%%%%%%%%%%%%%%%%%%%%%%%%%%%%%%%%%%%%%%%%%%%%%%

To analyse the partial null energy condition introduced earlier in this
paper, consider a generic null vector inclined at an angle $\psi$
away from the radial direction.  Then without loss of generality,
in an orthonormal frame attached to the $(t,r,\theta,\phi)$ coordinate
system,
\begin{equation}
k^{\hat \mu} \propto (\pm1,\cos\psi, 0, \sin\psi).
\end{equation}
Ignoring the (presently irrelevant) overall normalization of
the null vector, one has
\begin{eqnarray}
T_{\mu\nu} k^\mu k^\nu 
&\propto& (\rho -\tau \cos^2\psi + p \sin^2\psi) 
\nonumber\\
&=&
([\rho -\tau]  + [\tau +p ]\sin^2\psi).
\end{eqnarray}
We have already seen that $\rho-\tau$ is positive outside the event
horizon. On the other hand the Page approximation gives
\begin{equation}
\tau(r) +p(r) = - 24 \; p_\infty \;  \left({2M\over r}\right)^6,
\end{equation}
which is everywhere negative. A glance at Howard's numerical data
confirms that the numerical data also satisfies $\tau+p<0$.

\begin{figure}[htb]
\vbox{\hfil\epsfbox{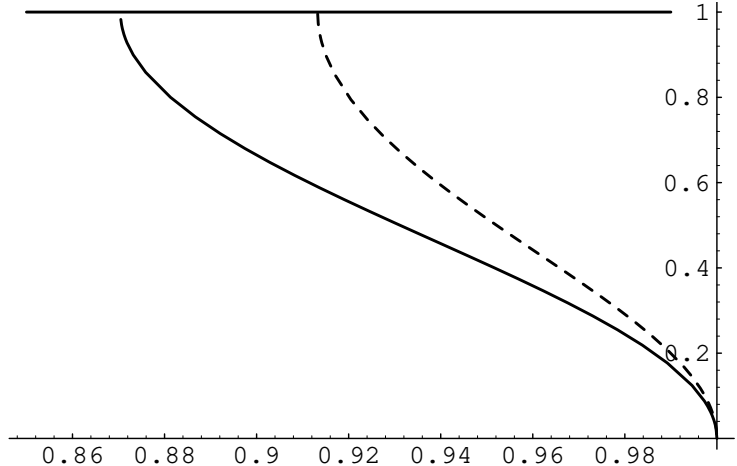}\hfil}
\caption[PNEC: Outside the horizon.]
{\label{fig1}
PNEC: Outside the horizon. This graph shows the (normalized) critical
angle $\psi_{\rm crit}(z)/[\pi/2]$ above which the partial null
energy condition fails. (The solid line represents the numerical
data; the dotted line represents the Page approximation.)}
\end{figure}

Thus the partial null energy condition {\em fails} for those radii
$r$, and those angles $\psi$, such that
\begin{equation}
\sin\psi > \sin[\psi_{\rm crit}(r)] \equiv  
\sqrt{ \rho(r)-\tau(r) \over |\tau(r) +p(r)| }.
\end{equation}
That is, the partial null energy condition {\em fails} for $\psi\in
[\psi_{\rm crit}(r),\pi/2]$. Defining $z=2M/r$ this critical angle
is (in the Page approximation) given by
\widetext
\begin{equation}
\psi_{\rm crit}(z) = 
\sin^{-1} \left[
\sqrt{ (1-z)
( 1 + 3z + 6z^2 + 10z^3 + 15z^4 + 21z^5 )/6z^6}\right].
\end{equation}
%%%\narrowtext
%%%%%%%%%%%%%%%%%%%%%%%%%%%%%%%%%%%%%%%%%%%%%%%%%%%%%%%%%%%%%%%%%%%%%%%%%%%%%%
%          5      4        3  2       2  3         4       5
%(r - R) (r  + 3 r  R + 6 r  R  + 10 r  R  + 15 r R  + 21 R )
%------------------------------------------------------------
%                               6
%                            6 R
%%%%%%%%%%%%%%%%%%%%%%%%%%%%%%%%%%%%%%%%%%%%%%%%%%%%%%%%%%%%%%%%%%%%%%%%%%%%%
For $r> 2.18994 M$ ($z< 0.913267$) there are no real solutions to
this equation.  At $r=2.18994 M$ one has $\psi_{\rm crit} = \pi/2$,
while $\psi_{\rm crit}(r)$ moves monotonically to zero as $r\to
2M$. (See figure~\ref{fig1} where $\psi_{\rm crit}(z)$ is plotted
as a function of $z$.)

Thus, sufficiently near the horizon, almost all directions violate
the partial null energy condition. As one moves further away from
the horizon the violations of the partial energy condition are
confined to null vectors that are progressively more and more
transverse, finally at $r=2.18994 M$ the violations of the partial
null energy condition disappear completely.

Using the numerical data will modify the precise location where
this behaviour manifests itself, but will not qualitatively modify
this picture. The violations of the partial null energy condition
vanish outside $r=2.298 M$ ($z=0.8703$), and the numerically
determined values of $\psi_{\rm crit}(z)$ are superimposed on
figure~\ref{fig1}.

%%%*** figure 1 should appear near here ***

%%%*******************************************************************
%%%*******************************************************************

%%%%%%%%%%%%%%%%%%%%%%%%%%%%%%%%%%%%%%%%%%%%%%%%%%%%%%%%%%%%%%%%%%%%%%%%%%%%%%
\subsection{Inside the horizon:}
%%%%%%%%%%%%%%%%%%%%%%%%%%%%%%%%%%%%%%%%%%%%%%%%%%%%%%%%%%%%%%%%%%%%%%%%%%%%%%

Inside the event horizon, one should consider a generic null vector
inclined at an angle $\tilde\psi$ away from the $t$ direction (which
is now spacelike). Then without loss of generality, in an orthonormal
frame attached to the $(t,r,\theta,\phi)$ coordinate system,
\begin{equation}
k^{\hat \mu} \propto (\cos\tilde\psi,\pm 1,0,\sin\tilde\psi).
\end{equation}
One should now consider the quantity
\begin{eqnarray}
T_{\mu\nu} k^\mu k^\nu 
&\propto&
(\tau -\rho\cos^2\tilde\psi + p\sin^2\tilde\psi)
\nonumber\\
&=& 
([\tau-\rho]  + [\rho +p ]\sin^2\tilde\psi).
\end{eqnarray}
Note that inside the horizon $\tau-\rho$ is strictly positive, while
$\rho+p$ is strictly negative. In fact, the Page approximation yields
\widetext
\begin{equation}
\rho(r) + p(r) =  4 p_\infty 
\left[1+ 2\left({2M\over r}\right)+ 3\left({2M\over r}\right)^2+ 
4\left({2M\over r}\right)^3+ 5\left({2M\over r}\right)^4+
6\left({2M\over r}\right)^5 - 27\left({2M\over r}\right)^6\right].
\end{equation}
%%%\narrowtext
%%%%%%%%%%%%%%%%%%%%%%%%%%%%%%%%%%%%%%%%%%%%%%%%%%%%%%%%%%%%%%%%%%%%%%%%%%%%%%%%%                
%                2      3      4      5       6
%4 (1 + 2 z + 3 z  + 4 z  + 5 z  + 6 z  - 27 z )
%%%%%%%%%%%%%%%%%%%%%%%%%%%%%%%%%%%%%%%%%%%%%%%%%%%%%%%%%%%%%%%%%%%%%%%%%%%%%%
One deduces that violations of the partial null energy condition
occur whenever
\begin{equation}
\sin\tilde\psi > \sin[\tilde\psi_{\rm crit}(r)] \equiv 
\sqrt{ \tau(r)-\rho(r) \over |\rho(r)+p(r)| }
\end{equation}
Consequently, violations of  the partial null energy condition are
confined to the range $\psi\in [\tilde\psi_{\rm crit}(r),\pi/2]$. Again
defining $z=2M/r$ this critical angle is (in the Page
approximation) given by
\widetext
\begin{equation}
\tilde\psi_{\rm crit}(z) = 
\sin^{-1} \left[
\sqrt{ 
{  (1-z) ( 1 + 3z + 6z^2 + 10z^3 + 15z^4 + 21z^5 ) 
\over
 (1 + 2 z + 3 z^2  + 4 z^3  + 5 z^4  + 6 z^5  - 27 z^6 ) }
}\right].
\end{equation}
%%%\narrowtext
%%%%%%%%%%%%%%%%%%%%%%%%%%%%%%%%%%%%%%%%%%%%%%%%%%%%%%%%%%%%%%%%%%%%%%%%%%%%%%
%                       2       3       4       5
%(-1 + z) (1 + 3 z + 6 z  + 10 z  + 15 z  + 21 z )
%-------------------------------------------------
%                2      3      4      5       6
%  -1 - 2 z - 3 z  - 4 z  - 5 z  - 6 z  + 27     %%%%%%%%%%%%%%%%%%%%%%%%%%%%%%%%%%%%%%%%%%%%%%%%%%%%%%%%%%%%%%%%%%%%%%%%%%%%%%
One has, at $r=0$, $\tilde\psi_{crit}(r=0) = \sin^{-1}\sqrt{7/9}
\approx 62\deg$, while $\tilde\psi_{\rm crit}$ falls to zero as
$r$ approaches $2M$.   Qualitatively: near (but inside) the horizon
almost all null directions suffer violations of the
partial energy condition, while near the central singularity somewhat
fewer directions violate the partial null energy condition.

Again, I remind the reader to not take the Page approximation too
seriously as one approaches the singularity.

%%% the old figure 2 has been deleted for brevity (See figure \ref{fig2}.)

%%%%%%%%%%%%%%%%%%%%%%%%%%%%%%%%%%%%%%%%%%%%%%%%%%%%%%%%%%%%%%%%%%%%%%%%%%%%%%
\section{Asymptotic null energy condition}
%%%%%%%%%%%%%%%%%%%%%%%%%%%%%%%%%%%%%%%%%%%%%%%%%%%%%%%%%%%%%%%%%%%%%%%%%%%%%%

I now turn attention to the asymptotic null energy condition
(Scri--NEC) that I introduced earlier in this paper. Discussing
this energy condition will require some standard results concerning
the null geodesics of the Schwarzschild spacetime. (See, for
instance, such standard textbooks as Misner--Thorne--Wheeler~\cite{MTW},
Wald~\cite{Wald}, or Weinberg~\cite{Weinberg}.)

%%%%%%%%%%%%%%%%%%%%%%%%%%%%%%%%%%%%%%%%%%%%%%%%%%%%%%%%%%%%%%%%%%%%%%%%%%%%%%
%%%\subsection{Outside the horizon:}
%%%%%%%%%%%%%%%%%%%%%%%%%%%%%%%%%%%%%%%%%%%%%%%%%%%%%%%%%%%%%%%%%%%%%%%%%%%%%%

There are two conserved quantities for null geodesic motion in
Schwarzschild spacetime; the energy and angular momentum. Using
these conservation laws the affine parameter can be chosen in such
a way that
\begin{eqnarray}
{dt\over d\lambda} &=& {1\over 1-2M/r};\\
{d\phi\over d\lambda} &=& {a^2\over r^2}.
\end{eqnarray}
The parameter $a$ is the angular momentum per unit energy. If the
null geodesic reaches asymptotic spatial infinity then this parameter
can also be interpreted as the ``impact parameter''. However, there
is a large class of null geodesics that never reaches spatial
infinity, for these null geodesics the notion of ``impact parameter''
is at best an abuse of language.

The angle between the null geodesic and the radial direction is
given by
\begin{equation}
\sin\psi = \sqrt{1-2M/r} \;\; \left({a\over r}\right).
\end{equation}
We are ultimately interested in the quantity
\begin{eqnarray}
T_{\mu\nu} k^\mu k^\nu 
&\propto&  
([\rho -\tau]  + [\tau +p ]\sin^2\psi)
\nonumber\\
&=&        
\{[\rho -\tau]  + [\tau +p ](1-2M/r)a^2/r^2\},
\end{eqnarray}
but will need to go through a few preliminaries.

The radial motion of null geodesics is governed by
\begin{equation}
\left({dr\over dt}\right)^2 = 
(1-2M/r)^2 \left[1-(1-2M/r){a^2\over r^2} \right].
\end{equation}
So the turning points, $dr/dt=0$, are given by the cubic
\begin{equation}
r^2 = a^2 (1-2M/r).
\end{equation}
If the impact parameter is small, $a<3\sqrt{3}M$, then it is a
standard result that there are no turning points: the null geodesic
either plunges into the future singularity, or emerges from the
past singularity of the maximally extended Schwarzschild spacetime.
(For $a\lesssim 3\sqrt{3}M$ the geodesic may make a large number
of ``orbits'' before crossing the event horizon.)

If the impact parameter is marginal, $a=3\sqrt{3}M$, then it is a
standard result that $r=3M$, corresponding to the (unstable) circular
photon orbit.

If the impact parameter is large, $a>3\sqrt{3}M$, then it is a
standard result that there are {\em two} turning points at physical
values of $r$. One of these turning points lies in the range
$r\in(3M,\infty)$, while the other lies in the range $r\in(2M,3M)$.

Note that if $a\gg 2M$ then the three mathematical roots of the cubic
are approximately $r\approx \pm a-M$, and $r\approx 2M[1+(2M/a)^2]$.
The two physical roots are approximately $r\approx a-M$ and $r\approx
2M[1+(2M/a)^2]$.

The first of these turning points, $r\in(3M,\infty)$, corresponds
to the obvious class of null geodesics with high impact parameter---those
geodesics that come in from spatial infinity, bounce off the angular
momentum barrier, and return to spatial infinity. (For $a\gtrsim
3\sqrt{3}M$ the geodesic may make a large number of ``orbits''
before returning to spatial infinity.)

The second of these turning points, $r\in(2M,3M)$, corresponds to
a completely separate class of null geodesics with high ``impact
parameter''---these geodesics emerge from the event horizon at
$t=-\infty$, with high angular momentum, make a large number of
``orbits'' before reaching their maximum height above the event
horizon, and then  make an equally large number of ``orbits'' before
returning to re-cross the event horizon at $t=+\infty$. For these
geodesics the use of the phrase ``impact parameter'' to describe
the parameter $a$ is most definitely an abuse of language.

To analyze the asymptotic null energy condition, start by first
considering the quantity
\begin{equation}
\Pi(r,a) \equiv (\rho -\tau \cos^2\psi + p \sin^2\psi),
\end{equation}
which I shall refer to as the NEC density.  (Remember that $\psi(r,a)$
is an explicitly known function of $r$ and $a$.)

Those null geodesics that come in from infinity and return to
infinity never get closer to the origin than $r=3M$. See, for
instance,~\cite[pages 672--678]{MTW}. Inspection of either the Page
approximation or of the numerical data indicates that $\rho$,
$-\tau$, and $p$ are all positive for $r \geq 3M$. Thus the asymptotic
null energy condition is satisfied along all null geodesics that
come from, and return to, infinity.

\begin{figure}[htb]
\vbox{\hfil\epsfbox{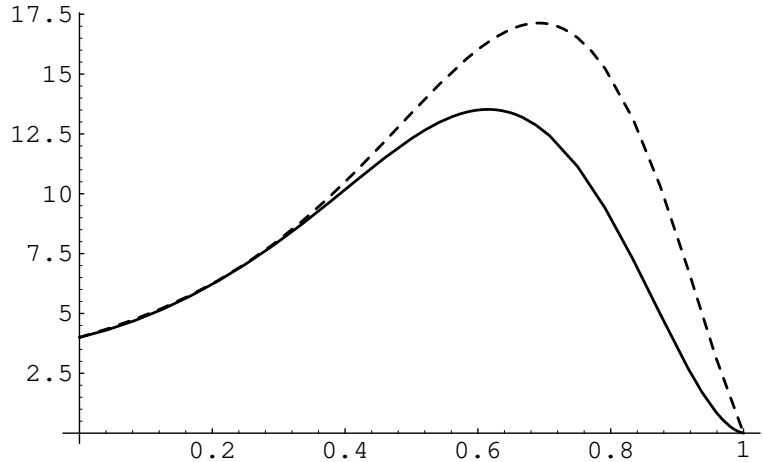}\hfil}
\caption[L(r): A bound on NEC violations.]
{\label{fig3}
L(z): A bound on NEC violations. For all impact parameters
$a<3\sqrt{3}M$, the NEC density $\Pi(r,a)$  is bounded below by
the quantity $L(r)$, which is itself bounded below by zero. This
implies that Scri--NEC is satisfied on all infalling or outfalling
null geodesics. (The solid line represents the numerical
data; the dotted line represents the Page approximation.)}
\end{figure}

For other classes of null geodesics it proves convenient to rewrite
the quantity $\Pi(r,a)$ as:
\begin{equation}
\Pi(r,a) \equiv \{[\rho -\tau]  + [\tau +p ](1-2M/r)a^2/r^2\}.
\end{equation}
Now, for incoming null geodesics with smaller than critical impact
parameter, the null geodesic may circle the black hole a large
number of times, but is guaranteed to ultimately plunge into the
event horizon~\cite[pages 672--678]{MTW}. This makes the analysis
a little more subtle. Inspection of either the Page approximation
or the numerical data shows that (outside the horizon) $\rho-\tau$
is always positive, while $\tau + p$ is always negative.  Now use
the fact that for an infalling null geodesic $a<3\sqrt{3}M$.  Since
$\tau + p$ is negative, this implies (for this class of null
geodesics) a lower bound
\begin{equation}
\Pi(r,a) > L(r) \equiv
\{[\rho -\tau]  + [\tau +p ](1-2M/r)27M^2/r^2\}.
\end{equation}
Inspection of Page's approximation indicates that  this lower bound
is strictly positive (zero at the horizon). In fact
\widetext
\begin{equation}
L(r) =  4p_\infty 
\left(1-{2M \over r}\right) 
\left[
1+ 3\left({2M\over r}\right)    +6\left({2M\over r}\right)^2 
+ 10\left({2M\over r}\right)^3 +15\left({2M\over r}\right)^4
+ 21\left({2M\over r}\right)^5 -{81\over2}\left({2M\over r}\right)^8
\right].
\end{equation}
%%%\narrowtext
%%%%%%%%%%%%%%%%%%%%%%%%%%%%%%%%%%%%%%%%%%%%%%%%%%%%%%%%%%%%%%%%%%%%%%%%%%%%%%
%              8      7         6  2       5  3       4  4       3  5       8
%2 (r - R) (2 r  + 6 r  R + 12 r  R  + 20 r  R  + 30 r  R  + 42 r  R  - 81 R )
%-----------------------------------------------------------------------------
%                                      9
%                                     r
%%%%%%%%%%%%%%%%%%%%%%%%%%%%%%%%%%%%%%%%%%%%%%%%%%%%%%%%%%%%%%%%%%%%%%%%%%%%%%
(Note the factorization!) Thus the Page approximation suggests that
the asymptotic null energy condition holds on all infalling null
geodesics. (At least until one crosses the horizon!) This lower
bound $L(r)$ is plotted in figure~\ref{fig3}.

By time reversal, this suggests that the asymptotic null energy
condition also holds on all outgoing null geodesics that reach
infinity.

%%%*** figure 2 near here ***

%%%****************************************************************
%%%****************************************************************

With respect to the numerical data, one can easily evaluate $L(r)$
numerically. The results are superimposed on figure~\ref{fig3}.
While there are marked differences between the analytic approximation
and the numerical data, both curves are seen to be strictly positive
outside the horizon.

We conclude that Scri--NEC is satisfied. For {\em all} null geodesics
that reach spatial infinity  the NEC density $\Pi(r,a)$ is strictly
positive everywhere outside the event horizon.

For completeness, and future reference, I point out that the Page
approximation yields
\widetext
\begin{equation}
\Pi(r,a) =  4p_\infty 
\left(1-{2M \over r}\right) 
\left[
1+ 3\left({2M\over r}\right)    +6\left({2M\over r}\right)^2 
+ 10\left({2M\over r}\right)^3 +15\left({2M\over r}\right)^4
+ 21\left({2M\over r}\right)^5 -{3a^2\over2}\left({2M\over r}\right)^8
\right].
\label{eq_Pi_r_a}
\end{equation}
Again, note the factorization!
%%%\narrowtext
%%%%%%%%%%%%%%%%%%%%%%%%%%%%%%%%%%%%%%%%%%%%%%%%%%%%%%%%%%%%%%%%%%%%%%%%%%%%%%
%%%                           2       3       4       5       2  8
%%%2 (-1 + z) (-2 - 6 z - 12 z  - 20 z  - 30 z  - 42 z  + 3 aa  z )
%%%%%%%%%%%%%%%%%%%%%%%%%%%%%%%%%%%%%%%%%%%%%%%%%%%%%%%%%%%%%%%%%%%%%%%%%%%%%%

%%%%%%%%%%%%%%%%%%%%%%%%%%%%%%%%%%%%%%%%%%%%%%%%%%%%%%%%%%%%%%%%%%%%%%%%%%%%%%
\section{Trapped null geodesics:}
%%%%%%%%%%%%%%%%%%%%%%%%%%%%%%%%%%%%%%%%%%%%%%%%%%%%%%%%%%%%%%%%%%%%%%%%%%%%%%

Now turn attention to the trapped null geodesics. (These are null
geodesics with high ``impact parameter'' $a>3\sqrt{3} M$, trapped
in the region $r\in[2M,3M]$.) While these trapped null geodesics
are, by definition, not directly relevant to the Scri--NEC, the
tools developed above permit us to gain additional insight into
the PNEC on trapped null geodesics.

Pick some value of $r$ in the range $(2M,3M)$. Then $\Pi(r,a)$ is
guaranteed to be negative if one chooses
\begin{eqnarray}
a > a_{\rm crit}(r) &\equiv& 
\sqrt{ {r^2 (\rho-\tau)\over(1-2M/r)|\tau+p|} }
\\
&=& 2M \sqrt{ {(\rho-\tau)\over z^2(1-z)|\tau+p|} }.
\end{eqnarray}
(Setting $z=2M/r$ is again a convenient choice of variables for
both analytic and numeric work. In this section we will only be
interested in the region $z\in[2/3,1]$.) Note that what we are
doing is guaranteeing that with these definitions the PNEC is
violated for the region of the $(a,z)$ plane above the curve $a_{\rm
crit}(z)$.

\begin{figure}[htb]
\vbox{\hfil\epsfbox{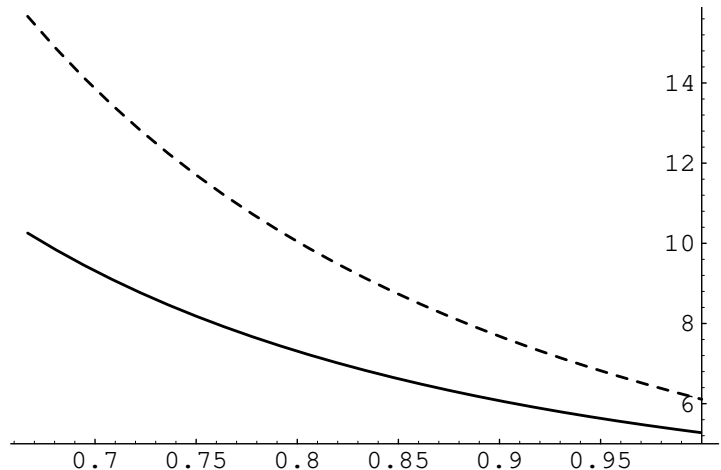}\hfil}
\caption[$a_{\rm crit}(r)$]
{\label{fig4}
$a_{\rm crit}(z)/M$: The critical impact parameter above which NEC
is violated on trapped null geodesics. (The solid line represents
the numerical data; the dotted line represents the Page approximation.)}
\end{figure}

Using Page's approximation, this critical
impact parameter is given by
\begin{equation}
a_{\rm crit}(z) = 2M 
\sqrt{(1 + 3z + 6 z^2 + 10 z^3 +15 z^4 + 21 z^5)/(6z^8) }.
\end{equation}
%%%%%%%%%%%%%%%%%%%%%%%%%%%%%%%%%%%%%%%%%%%%%%%%%%%%%%%%%%%%%%%%%%%%%%%%%%%%%%
%             2       3       4       5
%1 + 3 z + 6 z  + 10 z  + 15 z  + 21 z
%--------------------------------------
%                    8
%                 6 z
%%%%%%%%%%%%%%%%%%%%%%%%%%%%%%%%%%%%%%%%%%%%%%%%%%%%%%%%%%%%%%%%%%%%%%%%%%%%%%
%   3                  2      3      4    5
%-(u  (21 + 15 u + 10 u  + 6 u  + 3 u  + u ))
%--------------------------------------------
%                     6
%%%%%%%%%%%%%%%%%%%%%%%%%%%%%%%%%%%%%%%%%%%%%%%%%%%%%%%%%%%%%%%%%%%%%%%%%%%%%%
The critical impact parameter implied by the numerical data
was also determined and both curves are plotted on figure~\ref{fig4}.

%%%*** insert figure 3 near here ***

%%%*******************************************************************
%%%*******************************************************************

\begin{figure}[htb]
\vbox{\hfil\epsfbox{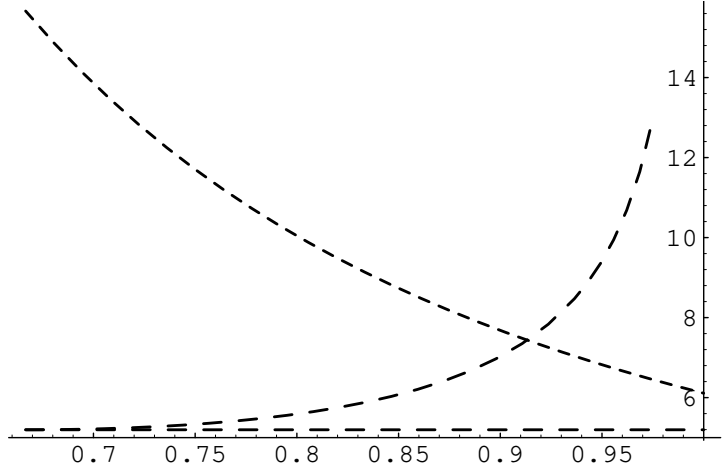}\hfil}
\caption[Page approximation: $a_{\rm crit}(r)$ and $a_{\rm max}(r)$]
{\label{fig5}
Page approximation: The critical impact parameter above which NEC
is violated ($a_{\rm crit}(z)/M$)is superimposed on the kinematic
bound ($a_{\rm max}(z)/M$). (The dotted line represents the Page
approximation; the dashed line represents the kinematic bound.) }
\end{figure}

\begin{figure}[htb]
\vbox{\hfil\epsfbox{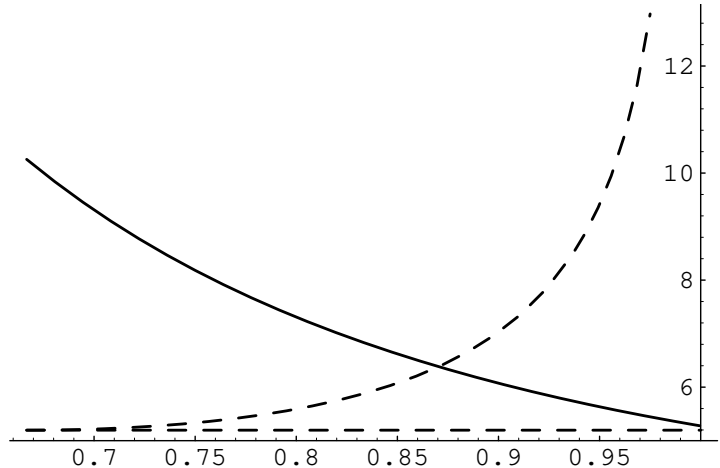}\hfil}
\caption[Numeric data: $a_{\rm crit}(r)$ and $a_{\rm max}(r)$]
{\label{fig6}
Page approximation: The critical impact parameter above which NEC
is violated ($a_{\rm crit}(z)/M$)is superimposed on the kinematic
bound ($a_{\rm max}(z)/M$). (The solid line represents the numerical
data; the dashed  line represents the kinematic bound.) }
\end{figure}

It should be noted that, for a given value of $r\in(2M,3M)$, one
cannot choose $a$ arbitrarily: In order for a trapped null geodesic
with impact parameter $a$ to ever reach radius $r$ it is necessary
that $a$ be small enough. Indeed one must have
\begin{equation}
a < a_{\rm max}(r) \equiv 
\sqrt{{r^2\over 1-2M/r}} = 
{2M\over \sqrt{z^2(1-z)}}.
\end{equation}
This may be thought of as a kinematic bound on trapped null geodesics:
the region of the $(a,z)$ plane above the curve $a_{\rm max}(z)$
is kinematically inaccessible to trapped null geodesics. (Note that
the region below $a=3\sqrt{3}M$ is also kinematically inaccessible.)
The relevant curve is plotted in figures~\ref{fig5} and~\ref{fig6},
where it is overlain with $a_{\rm crit}(z)$ as obtained from
figure~\ref{fig4}. By looking at where these curves cross one
another we can draw some general conclusions.

%%%*** insert figure 4 near here ***

%%%*******************************************************************
%%%*******************************************************************

In the Page approximation, now restricting attention to the region
outside the event horizon:
\begin{enumerate}
\item
Some trapped null geodesics (those with $a\in [3\sqrt{3}M, 2
\sqrt{7/3}M] = [5.19615 M, 6.1101 M]$) {\em never} experience NEC
density violations. That is: $\Pi(r,a)>0$ is {\em satisfied} along
the entire portion of the null geodesic that lies outside the event
horizon. (And therefore the [truncated] ANEC must be satisfied on
this entire class of null geodesics.)
\item
Some trapped null geodesics (those with $a>7.436 M$) {\em always}
experience NEC density violations. That is: $\Pi(r,a)<0$ along the
entire [truncated] null geodesic. (And therefore the [truncated]
ANEC must be violated on this entire class of null geodesics.)
\item
All other trapped null geodesics  (those with $a\in [6.1101 M,7.436
M]$) will experience some NEC violations when they get sufficiently
close to the event horizon. (And therefore investigating ANEC
violations on this class of null geodesics requires more work.)
\item 
Note that $a=6.1101 M$ corresponds to a null geodesic that reaches
a maximum radius $r_{\rm max}=2.18994 M$, a number that we have seen
before (when discussing the NEC and PNEC).
\end{enumerate}

%%%*** insert figure 5 near here ***

%%%*******************************************************************
%%%*******************************************************************

Use of the numerical data implies quantitative though not qualitative
changes.  We observe that
\begin{enumerate}
\item
The set of trapped null geodesics which {\em never} experience NEC density
violations is much smaller---those with $a\in [3\sqrt{3}M, 5.276
M] = [5.19615 M, 5.276 M]$. (The [truncated] ANEC must be satisfied
on this entire class of null geodesics.)
\item
Some trapped null geodesics (those with $a>6.383 M$) {\em always}
experience NEC density violations. (Therefore the [truncated] ANEC
must be violated on this entire class of null geodesics.)
\item
All other trapped null geodesics  (those with $a\in [5.276 M, 6.383
M]$) will experience some NEC violations when they get sufficiently
close to the event horizon. (And therefore investigating ANEC
violations on this class of null geodesics requires more work.)
\item 
Note that $a=6.383 M$ corresponds to null geodesic that reaches a
maximum radius $r_{\rm max}=2.298 M$, a number that we have seen before
(when discussing the PNEC).
\end{enumerate}

The message to be extracted is this: The numeric data and analytic
approximations are in good qualitative agreement with each other.
Null geodesics that reach null infinity are well behaved (Scri--NEC
is satisfied), but {\em some} of the trapped null geodesics encounter
NEC violations (and ANEC violations).

%%%%%%%%%%%%%%%%%%%%%%%%%%%%%%%%%%%%%%%%%%%%%%%%%%%%%%%%%%%%%%%%%%%%%%%%%%%%%%
\section{Averaged null energy condition}
%%%%%%%%%%%%%%%%%%%%%%%%%%%%%%%%%%%%%%%%%%%%%%%%%%%%%%%%%%%%%%%%%%%%%%%%%%%%%%

We are finally in a position to pin down precisely violations of
the ANEC itself. Although significant information can already be
extracted by using the point-wise energy conditions already discussed,
beyond a certain stage explicit use of the affine parameterization
must be invoked.  Write the ANEC integral as~\cite[page 117]{Visser}
\begin{eqnarray}
I_\gamma 
&\equiv&
\int_\gamma T_{\mu\nu} \; k^\mu \; k^\nu \; d\lambda,
\\
&=&
\int_\gamma (\rho -\tau \cos^2\psi + p \sin^2\psi) \; \xi^2 \; d\lambda,
\\
&=&
\int_\gamma ([\rho -\tau]  + [\tau +p ]\sin^2\psi) \; \xi^2 \; d\lambda,
\\
&=&
\int_\gamma \Pi(r,a) \; \xi^2 \; d\lambda.
\end{eqnarray}

Note that the {\em integrand} is proportional to the quantity
$\Pi(r,a)$ which has already been extensively discussed in the
context of the Scri--NEC. 

%%%%%%%%%%%%%%%%%%%%%%%%%%%%%%%%%%%%%%%%%%%%%%%%%%%%%%%%%%%%%%%%%%%%%%%%%%%%%%
\subsection{Scri--ANEC}
%%%%%%%%%%%%%%%%%%%%%%%%%%%%%%%%%%%%%%%%%%%%%%%%%%%%%%%%%%%%%%%%%%%%%%%%%%%%%%

The arguments previously adduced for the Scri--NEC can be
carried over wholesale to the Scri--ANEC. In particular:
\begin{itemize}
\item
For null geodesics that come in from spatial infinity and return
to spatial infinity $(a>3\sqrt{3}M;r>3M)$ the {\em integrand} is
everywhere positive and so the ANEC holds on this entire class of
geodesics. (Note that satisfaction of the ANEC does not arise from
the trivial observation that the heat bath contributes an asymptotically
constant and positive energy density far from the black hole---rather
one has the stronger statement that the integrand itself is positive
along the entire geodesic.)
\item
The fact that the integrand is positive implies that ANEC will also
be satisfied for curves with transverse smearing~\cite{Flanagan-Wald},
or curves with arbitrary positive weighting functions~\cite{Ford-Roman96}.
\item
For infalling or outfalling null geodesics that reach asymptotic
spatial infinity $(a<3\sqrt{3}M; r\in(2M,\infty))$ the integrand is
everywhere positive and ANEC is satisfied. [Caveat: I stop the ANEC
integral once it reaches the horizon.]
\item
This may be interpreted as follows: Because Scri--NEC is satisfied,
it automatically follows that Scri--ANEC is satisfied.
\item
For the unstable circular photon orbit $(a=3\sqrt{3}M; r=3M)$ the
integrand is everywhere positive and ANEC is satisfied.
\item
However, for trapped null geodesics $(a>3\sqrt{3}M;r\in[2M,3M])$
the integrand is no longer necessarily positive, and this is the
only case for which we will explicitly need to look at the ``weighting
function'' $\xi^2 d\lambda$ appearing in the ANEC integral.
\end{itemize}

%%%%%%%%%%%%%%%%%%%%%%%%%%%%%%%%%%%%%%%%%%%%%%%%%%%%%%%%%%%%%%%%%%%%%%%%%%%%%%
\subsection{ANEC on trapped null geodesics?}
%%%%%%%%%%%%%%%%%%%%%%%%%%%%%%%%%%%%%%%%%%%%%%%%%%%%%%%%%%%%%%%%%%%%%%%%%%%%%%

Observe that the ``weighting function'' appearing in the ANEC
satisfies~\cite[page 133, equations (12.60)--(12.63)]{Visser}
\widetext
\begin{equation}
\xi^2 d\lambda = 
g_{00} \left({dt\over d\lambda}\right)^2 d\lambda =
(1-2M/r)\left({1\over1-2M/r}\right)^2 (1-2M/r) dt = dt.
\end{equation}
%%%\narrowtext
Although this calculation was carried out for the Schwarzschild
geometry the result remains true for an arbitrary static spacetime---the
ANEC integral is simply the time-average along a null geodesic of
the local-Lorentz NEC integrand where the time average is to be
taken with respect to the natural static time coordinate.

For actual calculations it is much more practical to re-express
this as an integral with respect to the radial variable $r$ by
using
%\widetext
\begin{equation}
{dr\over\sqrt{1-2M/r}} = \cos\psi \; dt \; \sqrt{1-2M/r},
\end{equation}
%%%%\narrowtext
so that
\widetext
\begin{equation}
dt = 
{dr\over\cos\psi\; (1-2M/r)} =
{dr\over(1-2M/r) \; \sqrt{1-(1-2M/r)a^2/r^2}}.
\end{equation}
%%%\narrowtext 
Now this observation appears to weight the region near the event
horizon very heavily---because of the explicit pole at $r=2M$.
However, the integrand $\Pi(r,a)$ has a zero at the event horizon---in
the Page approximation one discovers an explicit factor of $(1-2M/r)$
[{\em Cf.} equation \ref{eq_Pi_r_a}], while the numerical data also
exhibits a first-order zero in $\Pi(r,a)$ at the horizon.  Thus
for trapped null geodesics one may write
\widetext
\begin{eqnarray}
I_\gamma(a) 
= 2 \int_{2M}^{r_{\rm max}(a)} 
{\Pi(r,a) \over (1-2M/r)}\; {dr \over \sqrt{1-(1-2M/r)a^2/r^2}},
\end{eqnarray}
%%%\narrowtext
\noindent and have some confidence that the integral actually converges at
the lower bound $r=2M$. As a penultimate step, recall that calculating
$r_{\rm max}(a)$ involves solving a cubic. It is more convenient
to parameterize the trapped geodesic by calculating the impact
parameter in terms of the maximum height attained by the null
geodesic:
\begin{equation}
a(r_{\rm max}) = {r_{\rm max}\over\sqrt{1-2M/r_{\rm max}}}.
\end{equation}
So for these trapped null geodesics one finds the ANEC integral is
\widetext
\begin{eqnarray}
I_\gamma(r_{\rm max}) 
= 2 \int_{2M}^{r_{\rm max}} 
{\Pi(r,r_{\rm max}) \over (1-2M/r)} \;
{dr\over\sqrt{1-[(1-2M/r)r_{\rm max}^2]/[(1-2M/r_{\rm max})r^2]}},
\end{eqnarray}
%%%\narrowtext
To actually evaluate this integral numerically it is useful to
change variables to $z=2M/r$, and $z_0 = 2M/r_{\rm max}$, with the
result that
\widetext
\begin{eqnarray}
I_\gamma(z_0) 
= 4M \int_{z_0}^{1} 
{\Pi(z,z_0) \over z^2(1-z)} \;
{dz \over \sqrt{1-[z^2(1-z)]/[z_0^2(1-z_0)]}},
\end{eqnarray}
%%%\narrowtext
With a little work, the square root can be seen to factorize
explicitly
\widetext
\begin{eqnarray}
I_\gamma(z_0) 
= 4M z_0 \sqrt{1-z_0} \; 
\int_{z_0}^{1} 
{\Pi(z,z_0) \over z^2(1-z)} \;
{1\over\sqrt{z^2+z_0^2+z z_0-z-z_0}} \;
{ dz \over \sqrt{z-z_0}} ,
\end{eqnarray}
%%%\narrowtext
This integral, though singular at the lower limit ($z=z_0$,
corresponding to $r=r_{\rm max}$), is now certainly finite.  While
Mathematica has resources to deal with $1/\sqrt{z}$ singularities
at the endpoints of the integration range, it reacts badly to such
singularities when the location is chosen dynamically. That is:
$\int_{z_0}^1 1/\sqrt{z-z_0}$ is handled badly. For this reason
the change of variables $z=z_0+w$, while being a formal mathematical
identity, leads to much better numerical behaviour for the integral:
\widetext
\begin{eqnarray}
I_\gamma(z_0) 
= 4M z_0\sqrt{1-z_0} \; 
\int_{0}^{1-z_0} 
{\Pi(z_0+w,z_0) \over (z_0+w)^2(1-w-z_0)} \;
{1\over\sqrt{(z_0+w)^2+z_0^2+(z_0+w)z_0-(z_0+w)-z_0}} \;
{ dw \over \sqrt{w} }.
\end{eqnarray}
%%%\narrowtext

\begin{figure}[htb]
\vbox{\hfil\epsfbox{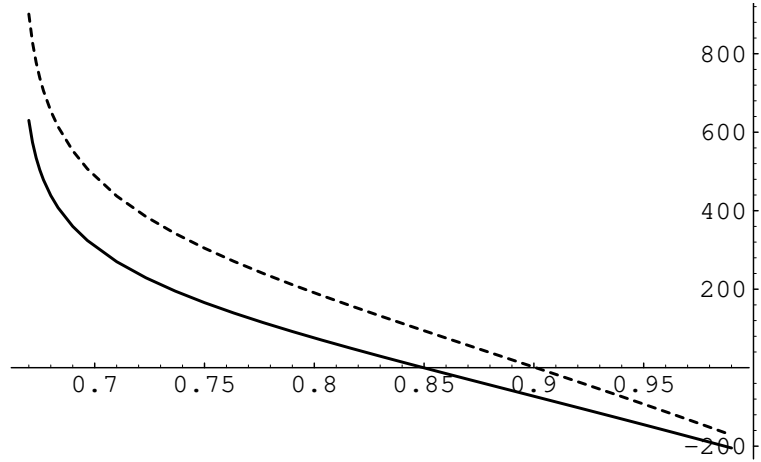}\hfil}
\caption[ANEC integral on trapped null geodesics.]
{\label{fig7}
ANEC integral on trapped null geodesics: The ANEC integral has been
evaluated numerically as a function of $z_0 = 2M/r_{\rm max}$. (The
solid line represents the numerical data; the dotted line represents
the Page approximation.) }
\end{figure}

This integral was determined numerically both for the Page
approximation and for the Anderson--Hiscock--Samuel numeric data.
The results are graphed in figure~\ref{fig7}. Remember that it is
meaningless to force this particular integral out of the range
$r\in[2M,3M]$, corresponding to $z_0\in[2/3,1]$

%%%*** insert figure 6 near here ***

%%%********************************
%%%********************************

Thus we have (finally!) managed to characterize the precise class
of null geodesics on which the [truncated] ANEC is violated.

The critical values of $z_0$, $r_{\rm max}$, and $a$ are, for the
Page approximation:
\begin{eqnarray}
z_0 &=& 0.90083;\nonumber\\
r_{\rm max} &=& 2.2202 \;M; \nonumber\\
a &=& 7.0501 \; M. 
\end{eqnarray}
For the numerical data, one obtains:
\begin{eqnarray}
z_0 &=& 0.8497;\nonumber\\
r_{\rm max} &=& 2.354 \; M; \nonumber\\
a &=& 6.071 \; M. 
\end{eqnarray}
Note that these numbers are qualitatively reasonable and in agreement
with the violations of the point-wise energy conditions, and the
numerical investigation of the PNEC. Furthermore the ANEC violations
in the numerical data are seen to extend out to larger distances
than the ANEC violations in the Page approximation, in agreement
with the general trend.

Though I have quoted four-significant-digit accuracy for the
numerical data one should probably not take anything past the second
significant digit too seriously. In fact, given the vagaries of
numerical analysis on singular integrals (and the inherent
uncertainties in the Anderson--Hiscock--Samuel data) it is conceivable
that the exact critical impact parameters might be $a=7M$ and $a=6M$
respectively.

%%%*** to be determined *** 

%%%%%%%%%%%%%%%%%%%%%%%%%%%%%%%%%%%%%%%%%%%%%%%%%%%%%%%%%%%%%%%%%%%%%%%%%%%%%%
\subsection{ANEC on non-geodesic curves?}
%%%%%%%%%%%%%%%%%%%%%%%%%%%%%%%%%%%%%%%%%%%%%%%%%%%%%%%%%%%%%%%%%%%%%%%%%%%%%%

If one considers arbitrary non-geodesic curves the ANEC loses much
(though not quite all) of its power: by a judicious choice of
non-geodesic null curve one could try to remain in a region where
the NEC is violated and thereby ``trivially'' violate the
ANEC~\cite{Klinkhammer91}.

For instance, along any circular null curve at fixed $r$ ({\em not}
a geodesic except in the case of $r=3M$)  the ANEC integral is
still proportional to $\rho+p$. Inspection of the numeric data
indicates that $\rho+p < 0$ for  $r \lesssim 2.298\, M$. (Page's
analytic approximation giving a slightly different result, $\rho+p
< 0$ for  $r < 2.18994\, M$.) Thus the ANEC is {\em not} satisfied
for this particular class of non--geodesic null curves.

%%%%%%%%%%%%%%%%%%%%%%%%%%%%%%%%%%%%%%%%%%%%%%%%%%%%%%%%%%%%%%%%%%%%%%%%%%%%%%
\section{Discussion}
%%%%%%%%%%%%%%%%%%%%%%%%%%%%%%%%%%%%%%%%%%%%%%%%%%%%%%%%%%%%%%%%%%%%%%%%%%%%%%

Investigation of the properties of the averaged null energy condition
is of considerable interest to diverse applications in semiclassical
quantum gravity. It is now abundantly clear that, in the test-field
limit, semiclassical quantum fields do {\em not} generally satisfy
the ANEC: Indeed ANEC violations are related to the existence of
a non-zero scale anomaly~\cite{Visser,Visser95}. Even if the scale
anomaly vanishes, this does not necessarily imply that the ANEC is
satisfied:  one has to do a case by case analysis.  As an example,
this paper investigates the situation in Schwarzschild spacetime.

The analysis presented herein is somewhat of an attempt to crack
a walnut with a sledgehammer in the sense that it is a collection
of rather general techniques applied to a rather particular
problem---it is certainly true that this type of analysis can now
be carried forward to other geometries, other quantum fields, and
other vacuum states by by straightforward but tedious computation.

For the Schwarzschild geometry (with a conformally coupled scalar
field in the Hartle--Hawking vacuum) the results may be expressed
thusly:
\begin{itemize}
\item Point-wise energy conditions---
\begin{itemize}
\item
Inside the event horizon, with suitable caveats regarding the
applicability of the Page approximation, almost any energy condition
you can think of will be violated.
\item
Between the event horizon $(r=2M)$ and the unstable photon orbit
$(r=3M)$ many of the energy conditions are violated, in a
series of onion-like layers.
\item
Outside the unstable photon orbit $(r=3M)$ all energy conditions
are satisfied.
\item
All null geodesics that reach asymptotic infinity are well-behaved:
If you look {\em along} the null geodesic you never see NEC violations.
\end{itemize}
\item Averaged null energy condition---
\begin{itemize}
\item
If your null geodesic ever reaches infinity, the [truncated] ANEC
is definitely satisfied, and is satisfied for a non-trivial reason:
the {\em integrand} is strictly positive all the way from the event
horizon to null infinity.
\item
Between the event horizon $(r=2M)$ and the unstable photon orbit
$(r=3M)$ some of the trapped null geodesics violate the [truncated]
ANEC.
\item
If you are willing to look at non-geodesic null curves even more
violations of the ANEC can be found.
\end{itemize}
\end{itemize}

It should be possible to generalize the observations of this paper.
For instance:
\begin{enumerate}
\item 
It would be very nice to have an analytic understanding of the
precise role played by the unstable photon orbit---the numerical
evidence is suggestive but not definitive.
\item
For that matter it would be nice to know if the relevance of the
unstable photon orbit generalizes to other geometries.
\item
It would be nice to go beyond the numerics; to develop some exact
analytic arguments that go beyond the Page approximation.
\item 
Generalizations to the Boulware vacuum will be presented in a
companion paper.
\item
Generalizations to the Unruh vacuum, other quantum fields,
non-conformal couplings, particle masses, and the Reissner--Nordstr\"om
geometry will be straightforward if tedious.
\item
The new energy conditions I introduce, Scri--NEC and Scri--ANEC,
are interesting in that they focus attention on null infinity. And
null infinity is where all the interesting details arise in the
Friedman--Schleich--Witt~\cite{Topological-censorship} topological
censorship theorem, and the Penrose--Sorkin--Woolgar version of
the positive mass theorem~\cite{Penrose-Sorkin-Woolgar}. I suspect
that with a little more work suitable generalizations of these
theorems can be constructed in terms of the Scri--ANEC.
\item
Finally I should point out that even though the various energy
conditions are violated in many regions, this does not give one a
completely free hand to design spacetime geometries to taste: It
seems quite likely that the ``quantum inequality'' approach of Ford
and Roman~\cite{Ford-Roman94,Ford-Roman95b} will allow us to place
constraints on spacetime geometries even if all the usual types of
energy condition fail.
\end{enumerate}

%%%%%%%%%%%%%%%%%%%%%%%%%%%%%%%%%%%%%%%%%%%%%%%%%%%%%%%%%%%%%%%%%%%%%%%%%%%%%%
\acknowledgements
%%%%%%%%%%%%%%%%%%%%%%%%%%%%%%%%%%%%%%%%%%%%%%%%%%%%%%%%%%%%%%%%%%%%%%%%%%%%%%

I wish to thank Paul Anderson for kindly making available
machine-readable tables of the numeric data used in this analysis.

I also wish to thank Nils Andersson, Paul Anderson, \'Eanna Flanagan,
Larry Ford, and Tom Roman for their comments and advice.

The numerical analysis in this paper was carried out with the aid
of the Mathematica symbolic manipulation package.

This research was supported by the U.S. Department of Energy.

%%%%%%%%%%%%%%%%%%%%%%%%%%%%%%%%%%%%%%%%%%%%%%%%%%%%%%%%%%%%%%%%%%%%%%%%%%%%%%
%%%%%%%%%%%%%%%%%%%%%%%%%%%%%%%%%%%%%%%%%%%%%%%%%%%%%%%%%%%%%%%%%%%%%%%%%%%%%%

%%%%%%%%%%%%%%%%%%%%%%%%%%%%%%%%%%%%%%%%%%%%%%%%%%%%%%%%%%%%%%%%%%%%%%%%%%%%%%
\end{document}